\documentstyle[prd,aps,floats,twocolumn,epsfig]{revtex}
\begin{document}

\title{\rightline{\small FUB-HEP/97-13}
       \vspace*{-1mm}
       \rightline{\small CPHT-S591.1297}
  \vskip 0.4cm
  High-statistics finite size scaling analysis of U(1) lattice
  gauge theory with Wilson action}
\author{Burkhard Klaus}
\address{Institut f\"ur Theoretische Physik,
  Freie Universit\"at Berlin, D-14195 Berlin, Germany\\
  and DESY-IfH Zeuthen, D-15735 Zeuthen, Germany}
\author{Claude~Roiesnel}
\address{Centre de Physique Th\'eorique,
  Centre National de la Recherche Scientifique, UPR A0014,\\
  Ecole Polytechnique, 91128 Palaiseau Cedex, France}
\date{December 1997}

\maketitle

\begin{abstract} 
We describe the results of a systematic high-statistics Monte-Carlo
study of finite-size effects at the phase transition of compact U(1)
lattice gauge theory with Wilson action on a hypercubic lattice with
periodic boundary conditions. We find unambiguously that the critical
exponent $\nu$ is lattice-size dependent for volumes ranging from
$4^{4}$ to $12^{4}$.  Asymptotic scaling formulas yield values
decreasing from $\nu(L\geq 4)\approx 0.33$ to $\nu(L\geq 9)\approx
0.29$. Our statistics are sufficient to allow the study of different
phenomenological scenarios for the corrections to asymptotic scaling.
We find evidence that corrections to a first-order transition with
$\nu=0.25$ provide the most accurate description of the data. However
the corrections do not follow always the expected first-order
pattern of a series expansion in the inverse lattice volume $V^{-1}$.
Reaching the asymptotic regime will require lattice sizes greater than
$L=12$.  Our conclusions are supported by the study of many
cumulants which all yield consistent results after proper
interpretation.
\end{abstract}

\section{Introduction}

The U(1) lattice gauge model has been studied from the early days of
numerical simulations of gauge theories \cite{CJR}. Its apparent
simplicity makes it a natural choice for testing concepts and
techniques used in simulations of non-abelian lattice gauge theories.
However it turned out that the abelian character of the U(1) model
brings specific difficulties. It can be proven rigorously under rather
general assumptions \cite{FRO} that abelian lattice gauge theories
have a phase transition at finite coupling though the order is not
known. The determination by numerical simulations of the properties of
the critical point in compact U(1) lattice gauge theory with Wilson
action and periodic boundary conditions has been strongly dependent of
the computing power available at the time.

The early simulations \cite{LN80} made runs with a few hundred
iterations at each coupling constant on rather small ($4^{4}$ to
$6^{4}$) lattices.  They reported a second-order phase transition but
observed some signs of metastability with longer runs at isolated
points. Subsequent simulations \cite{JNZa} made runs with a few
thousand iterations on larger ($8^{4}$ up to $16^{4}$) lattices.  It
became possible to histogram the runs and to observe a gap in the
plaquette energy. They concluded to the existence of a first-order
critical point but did notice a decrease of the gap with the lattice
size which was later confirmed \cite{GNC} on $9^{4}$ and
$(3\sqrt{3})^{4}$ lattices.

But the situation became still more confusing in the following years.
Indeed the results, always for lattices with periodic boundary
conditions and with runs of $O(10^{4})$ length, have usually depended
on which method was used for studying the phase transition. Most
studies using the Monte-Carlo renormalization group transformation
\cite{BUR,LAN,LR,DHH} have claimed a second order phase transition, but
sometimes with very different critical exponents, whereas most works
based on finite size scaling theory \cite{EJNZ,JNZb,BLSU} have
concluded to a first-order transition, but again with a critical
exponent not consistent with the first-order prediction. These
discrepancies led the subject into an expectant state.

The interest about the nature of the phase transition has been revived
more recently by two new kinds of results. The role played by
topological excitations, the monopoles, had been the object of
inquiries since the very beginning of numerical studies of the U(1)
lattice gauge theory \cite{DGT}. The influence of topology on the
critical properties can be probed in several ways. One approach
\cite{BAR,BSS,BS,KRWa} is to add to the standard Wilson action a
coupling $\lambda$ controlling the density of monopoles. There is some
evidence \cite{KRWb} in favor of the existence of a non-gaussian
second-order critical point in the $(\beta,\lambda)$ plane when
monopoles are suppressed. There is also preliminary evidence \cite{KD}
against universality with respect to $\lambda$. Another approach
\cite{LN94,JLNa} is to study the compact U(1) gauge theory on lattices
with sphere-like topology with a Wilson action extended by a coupling
$\gamma$ of charge 2
\begin{equation}
  S=\beta \sum_{P} \cos \Theta_P + \gamma \sum_{P} \cos 2\Theta_P
\end{equation}
No signs of metastability are found on these lattices for $\gamma \leq
0$. A finite-size scaling analysis of the data on sphere-like
lattices \cite{JLNb} has concluded to the existence, for $\gamma \leq
0$, of a second-order transition with a non-gaussian continuum limit.
However we must notice that the critical exponents of these two
approaches are still different.

The latter result has spawned further investigations. There has been a
study \cite{COX} of the scaling behaviour of gauge-ball masses and of
the static potential, which claims to confirm the second-order nature
of the transition also on lattices with periodic boundary conditions
at $\gamma = -0.2$ and $\gamma = -0.5$ with a critical exponent
$\nu\approx 0.36$ consistent with universality and the finite-size
scaling analysis of sphere-like lattices. The results on lattices with
spherical topology could also lend support to the suspicion that the
first-order signal observed on toroidal lattices might be a
finite-size effect due to the topology of the lattice. This suspicion
had been raised by the observation of the decrease of the gap
mentioned above. However the measurements of gaps were not accurate
enough and systematic enough to make a statement about the infinite
volume limit.  The particular case of the Wilson action was therefore
re-investigated \cite{KLA,ROI} with increased statistics ($O(10^{5})$
for each coupling constant) and an emphasis on the study of gaps.
There was definite evidence that the latent heat extrapolates to a
non-zero value in the infinite volume limit of toroidal lattices.
Moreover the measurement of the critical exponent, $\nu\approx 0.33$,
though clearly different from the first-order prediction, was not
statistically consistent with the claim of universality with respect
to the coupling $\gamma$.

Therefore there is always an apparent contradiction between the
simulations on lattices with sphere-like topology and on lattices with
periodic boundary conditions since one observes a gap on the latter
even when $\gamma<0$.  Moreover the dispersion of the predicted values
for the critical exponent $\nu$ is always as large as before, despite
the increase in statistics since the last decade. One can fairly state
that the confusion about the nature of the transition has also
increased! But one can now strongly suspect that the origin of these
paradoxical results could be explained by the presence of systematic
corrections to the asymptotic finite-size scaling formulas. This will
be our working hypothesis. There are two ways of circumventing these
corrections. The most straightforward approach is to simulate the
largest possible lattices for a given computing power. However, for a
transition with a weak first-order signal like U(1), one must
thoroughly assess whether the number of tunnelling events is
sufficient to get statistically reliable results.  The other approach
is to perform a finite-size scaling analysis of the four-dimensional
compact U(1) gauge theory on smaller lattices but with the same
quality standard as the best analyses of three-dimensional spin
models. Such an analysis must meet two criteria. On the one hand the
simulation must be done on many lattice sizes (and not restricted to 3
or 4 data points as is so often the case) in order to be able to test the
{\it stability} of the fits. On the other hand the statistics must be
large enough to {\it disentangle} the systematic errors from the
statistical errors. Only then can we interpret correctly the observed
deviations in a critical exponent extracted from different
observables. We can learn from the two and three dimensional numerical
studies of spin models how large the statistics must be: one needs at
least $O(10^{6})$ configurations at each coupling constant.  None of
the presently published studies of the U(1) phase transition satisfy
both constraints and we contend that this is the source of the
contradictory results.

The purpose of this work is to present a study at eight lattice sizes
with statistics up to $O(10^{7})$ at each pseudo-critical coupling
after reweighting and to clarify the nature of the U(1) phase
transition. In the next section we shall review the part of
finite-size scaling theory that we shall need in the sequel to
interpret the data. Then we shall describe the details of our
Monte-Carlo simulation and present the results of the measurements. In
the following sections we shall discuss, for various cumulants, the
finite-size scaling analysis of their pseudo-critical couplings and of
their extrema. The conclusion will be devoted to putting all these
results into a coherent perspective.

\section{Framework of the finite-size scaling analysis}
\label{frame}

\subsection{Scaling ansatz}

Most of the numerical finite-size scaling analyses of the phase
transition in the U(1) lattice gauge theory have been restricted to
the specific-heat and the Binder cumulant. When scaling violations are
important, as it turns out for U(1), introducing higher-order
cumulants brings some improvement as we shall explain below.  Most
reviews on finite-size scaling theory discuss the magnetic cumulants
only.  Studying energy cumulants carries some specific features.  Even
if this is standard lore we shall briefly remind some useful formulas
not always easy to find elsewhere.  First, to fix our notations, we
recall that the canonical partition function of a lattice gauge theory
on a $d$-dimensional cubic lattice of size $L$ can be written as
\begin{eqnarray}
  Z(\beta, L) & = & \int \Omega(E,L)\,e^{-\beta\,V\,E}\,dE \\
              & = & e^{-V F(\beta,L)}
\end{eqnarray}
where $\Omega(E,L)$ is the microcanonical partition function, $E$ is
the plaquette energy and the volume, for a lattice gauge theory, is
$V= \frac{1}{2} d(d-1)L^{d}$.

Standard arguments \cite{FIS} lead for continuous phase transitions
to the free energy density decomposition in a singular part and an
analytic contribution:
\begin{eqnarray}
  F(\beta,L)=L^{-d}f_{0}(x)+f_{ns}(t,L)
\end{eqnarray}
where $x=|t| L^{\frac{1}{\nu}}$ is the scaling variable and
$t=\frac{\beta}{\beta_{c}}-1$ is the reduced coupling. In this
parameterization, the hyperscaling relation $\alpha=2-\nu d$ is assumed
with only one relevant scaling variable $x$ and no dangerous
irrelevant variable. Moreover one assumes that there are no marginal
variables and no logarithmic bulk singularities, even if we work in
dimension $d=4$, which could be the upper critical dimension of the
continuum limit if there were a second-order phase transition.

The function $f_{ns}$ represents the non-singular contribution of the
background to the free energy density. $f_{ns}(t,L)$ is an analytic
function in $t$. Its dependence upon $L$ is not so clear.  There is a
conjecture \cite{PHA} that for periodic systems one can take
$f_{ns}(t,L) \approx f(t,\infty)$. We shall assume that this conjecture
holds true and that the $L$-dependence is exponentially suppressed:
\begin{eqnarray}
  f_{ns}(t,L)=f_{00}+f_{01}t+f_{02}t^{2}+\cdots+O\left(e^{-L/\xi}\right)
\end{eqnarray}
However one must be aware that for systems with free boundary
conditions one expects in general that $F(\beta,L)$ can have
geometrical terms present, which are proportional to inverse integral
powers of $1/L$.

There are further analytic corrections introduced by non-linearities
in the scaling variables away from criticality either when solving the
renormalization group beyond the linear approximation,
$t'=t+O(t^{2})$, or for instance, when using $t'=\beta_{c}/\beta-1$
instead of $t$. Expressing the non-linear scaling variable
$x'=t'L^{1/\nu}$ in terms of $x$ yields corrections in powers of
$L^{-\frac{1}{\nu}}$. We will neglect these corrections.

In the renormalization group description of second-order phase
transitions, one expects corrections to the asymptotic scaling
function $f_{0}(x)$ induced by the irrelevant variables. We shall keep
the leading contribution in these corrections and parameterize it by an
exponent $\omega>0$. Therefore we write
\begin{eqnarray}
  \label{ansatz}
  F(\beta,L)=L^{-d} \left(f_{0}(x)+L^{-\omega}f_{1}(x)\right)+f_{ns}(t)
\end{eqnarray}
We stress that this ansatz for the free energy density is a natural,
but completely phenomenological generalization of the asymptotic
scaling formula. Moreover this ansatz still depends at least upon 4
parameters. It is not yet possible to extract that many parameters from
a numerical finite-size analysis. We will have to resort to further
approximations and limit ourselves to considering three-parameter
fitting ans\"atze.

Ansatz~(\ref{ansatz}) can also describe the finite-size behavior at
first-order transitions. Even if there is not so much known about this
behavior from a theoretical standpoint, one usually expects \cite{FB}
that $\nu^{-1}=\omega=d$. Heuristic arguments based on the
double-gaussian approximation \cite{CLB,B90,LK} predict that the
corrections should be expressible as a power series in $V^{-1}$. These
arguments can be put onto a rigorous basis \cite{BK,BKM,BNB} in the
special case of q-states Potts models for large q.

\subsection{Standard cumulants}

In principle we would better like to determine directly the
finite-size scaling properties of $\Omega(E,L)$ or equivalently of the
probability distribution $P(E,L)$ of the plaquette energy. In practice
it is more convenient to extract the different moments of the
plaquette distribution:
\begin{eqnarray}
  \left<E^{n}\right>    & = &%
  \frac{(-1)^{n}}{V^{n}}\frac{1}{Z(\beta,L)}\frac{\partial^{n} Z}%
  {\partial\beta^{n}} \\
  \frac{\partial\left<E^{n}\right>}{\partial\beta} & = &%
  V\left(\left<E\right>\left<E^{n}\right>-\left<E^{n+1}\right>\right)
\end{eqnarray}
Then, with ansatz (\ref{ansatz}) it is easy to derive the scaling
properties of any cumulant from these moments.  The choice of
cumulants to include in the analysis is, in a large respect, rather
arbitrary.  Of course, for the sake of comparison with previous works,
we have to study standard cumulants such as the specific heat:
\begin{eqnarray}
  C_{v}(\beta,L) & = &%
  -\beta^{2}\frac{\partial^{2}}{\partial\beta^{2}} F(\beta,L)
  \\ & = &%
  \beta^{2}\,V\left(\left<E^{2}\right> - \left<E\right>^{2}\right)
\end{eqnarray}
and the Binder cumulant:
\begin{eqnarray}
  U_{4}(\beta,L) & = &%
  \frac{1}{3}\left(1-\frac{\left<E^{4}\right>}{\left<E^{2}\right>^{2}}\right)
\end{eqnarray}
We shall also introduce the second cumulant: 
\begin{eqnarray}
  U_{2}(\beta,L) & = &%
  1 - \frac{\left<E^{2}\right>}{\left<E\right>^{2}}
\end{eqnarray}
These low-order cumulants are sensitive to the analytic contribution
to the free energy density in Eq.~(\ref{ansatz}). For instance
ansatz~(\ref{ansatz}) implies the following for the scaling behavior
of the specific heat:
\begin{eqnarray}
  \label{Cvfss}
  C_{v}(\beta,L) & = &%
  - \frac{\beta^{2}}{\beta_{c}^{2}} L^{-d+\frac{2}{\nu}} \left( f''_{0}(x)%
  + L^{-\omega} f''_{1}(x) \right) - 2 f_{00}\frac{\beta^{2}}{\beta_{c}^{2}}  
\end{eqnarray}
But we know that $\frac{1}{d} \leq \nu \leq \frac{1}{2}$ and we can
expect that $\omega \approx \frac{2}{\nu}-d$. Such an approximate
equality would imply that the next-to-leading contribution comes both from
the corrections to scaling and from the analytic background. Then we
can anticipate difficulties in describing the scaling behavior of
the specific heat with three-parameter ans\"atze.

In the same way the predicted finite-size scaling behavior of the
second cumulant and the Binder cumulant is:
\begin{eqnarray}
  \label{U2fss}
  U_{2}(\beta,L) & = &%
  L^{-2d+\frac{2}{\nu}}\biggl(u_{20}(x) + u_{21}(x) L^{-d+\frac{1}{\nu}}
  \nonumber\\ & & \qquad +\,
  O\left(L^{d-\frac{2}{\nu}},L^{-2d+\frac{2}{\nu}},L^{-\omega}\right)\biggr)
\end{eqnarray}
\begin{eqnarray}
  \label{U4fss}
  U_{4}(\beta,L) & = &%
  L^{-2(d-\frac{1}{\nu})}\biggl(u_{40}(x) + u_{41}(x)L^{-(d-\frac{1}{\nu})}
  \nonumber\\ & & \qquad +\, 
  O\left(L^{-2(d-\frac{1}{\nu})},L^{d-\frac{2}{\nu}},L^{-\omega}\right)\biggr)
\end{eqnarray}
The corrections to scaling are now governed by the exponent
$d-\frac{1}{\nu}$. We can expect to run into problems describing
these corrections when $\nu \approx \frac{1}{d}$. In the limit $\nu =
\frac{1}{d}$, for first-order transitions,
Eqs.~(\ref{Cvfss},\ref{U2fss},\ref{U4fss}) yield the correct leading
behavior but the corrections become of order $L^{-d}$.

One can notice \cite{BIN} that deriving
Eqs.~(\ref{Cvfss},\ref{U2fss},\ref{U4fss}) with respect to $\beta$
introduces an additional factor $L^{\frac{1}{\nu}}$ through the
scaling variable $x$. Therefore studying the finite-size scaling
behavior of the derivatives of the standard cumulants should make it
easier to determine the critical exponent $\nu$. We have also
introduced these cumulants into our analysis:

\begin{eqnarray}
\frac{\partial C_{v}}{\partial\beta} & = &%
V\biggl[2\beta\left(\left<E^{2}\right>-\left<E\right>^{2}\right)
\nonumber\\ & & \qquad -\,
\beta^{2}V\left(\left<E^{3}\right>-3\left<E^{2}\right>\left<E\right>%
+2\left<E\right>^{3}\right)\biggr]
\end{eqnarray}
\begin{eqnarray}
\frac{\partial U_{2}}{\partial\beta} & = &%
\frac{V}{\left<E\right>^{3}}\left(\left<E^{3}\right>\left<E\right>%
+\left<E^{2}\right>\left<E\right>^{2}-2\left<E^{2}\right>^{2}\right)
\end{eqnarray}
\begin{eqnarray}
\frac{\partial U_{4}}{\partial\beta} & = &%
\frac{V}{3\left<E^{2}\right>^{3}}\Bigl(\left<E^{5}\right>\left<E^{2}\right>%
+\left<E^{4}\right>\left<E^{2}\right>\left<E\right>
\nonumber\\ & & \qquad\qquad -\,
2\left<E^{4}\right>\left<E^{3}\right>\Bigr)
\end{eqnarray}

\begin{figure}
\center
  \begin{tabular}{c}
  \epsfig{file=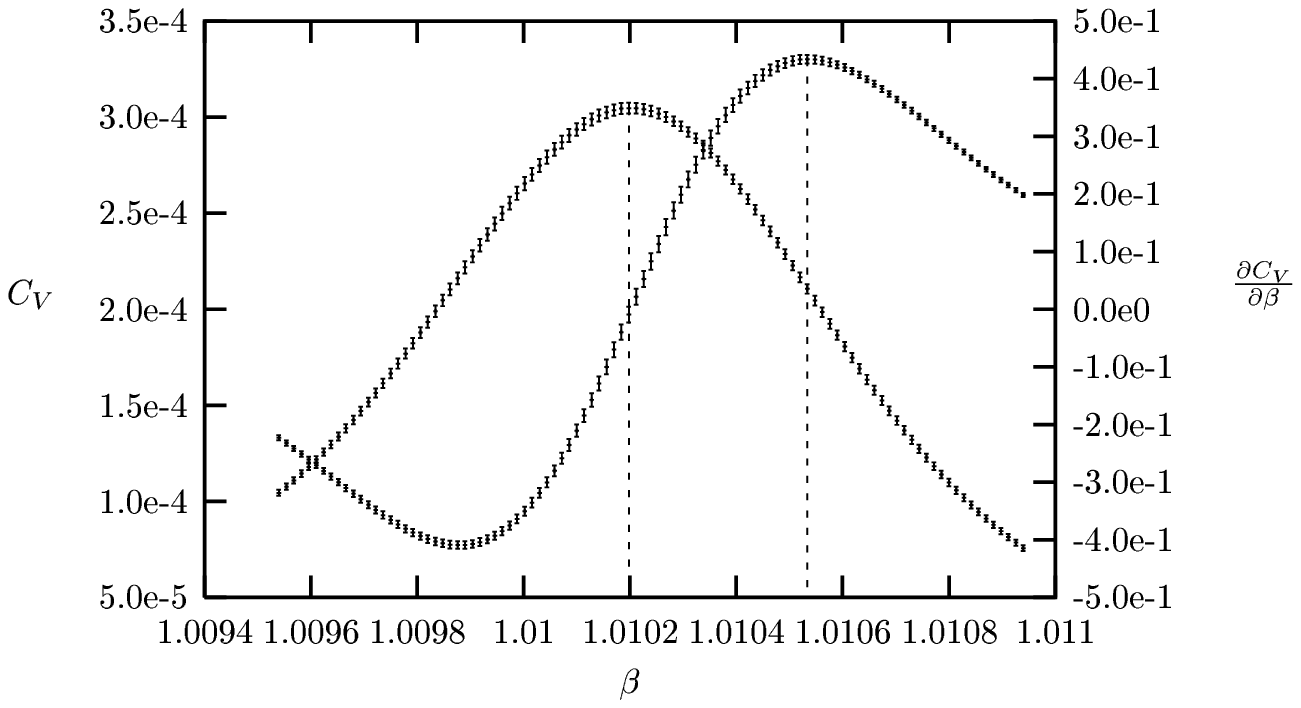, width=\linewidth} \\
  \epsfig{file=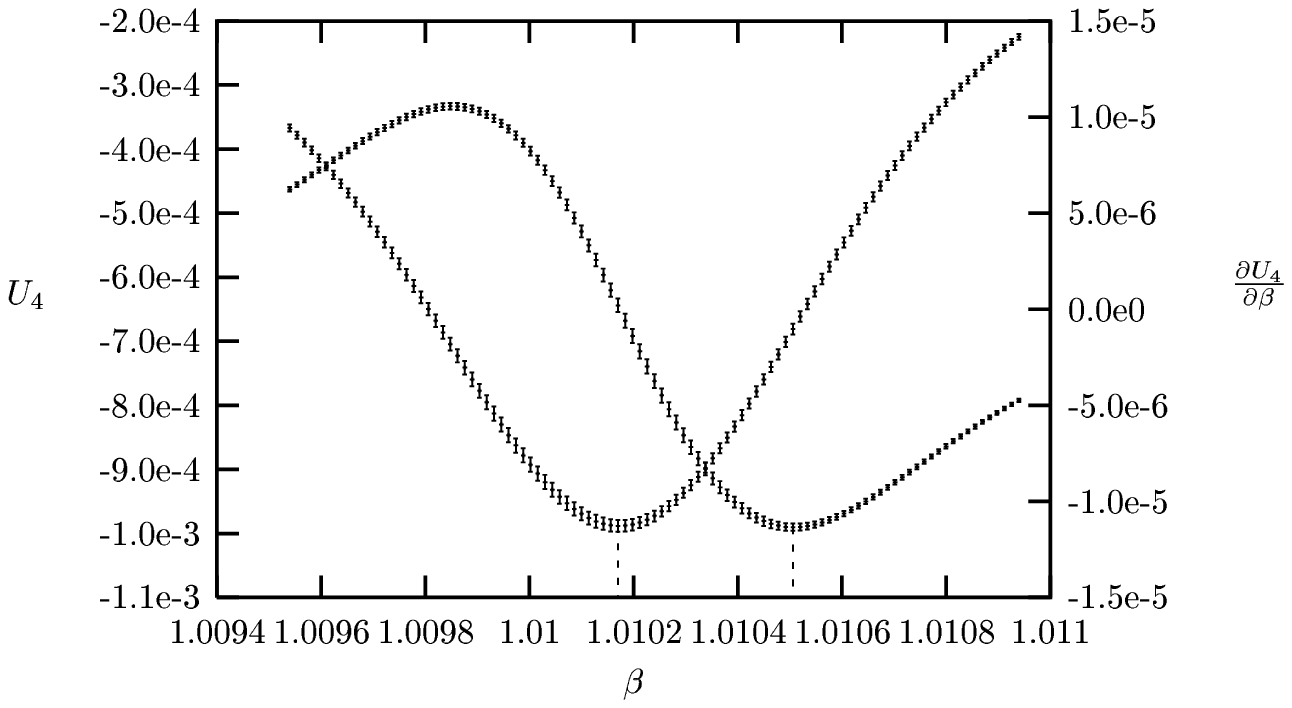, width=\linewidth}
  \end{tabular}
  \caption{The cumulants $C_{v}\times V^{-1}, U_{4}$ and their derivatives
           $\frac{\partial C_{v}}{\partial\beta}\times V^{-1},
           \frac{\partial U_{4}}{\partial\beta}\times V^{-1}$ on a 
           $12^4$ lattice as a function of $\beta$.}
  \label{fig:std}
\end{figure}

In order to make a comparison between different lattice sizes one
still needs a prescription to define a common value of the scaling
variable $x$. One usually uses the location of an extremum of a
cumulant $\kappa(\beta,L)$ to define a universal value
$x_{\kappa}^{*}$, independent of $L$. When there are several extrema
it is natural to choose the extremum closest to the infinite volume
critical point. Fig.~\ref{fig:std} displays the plots of the cumulants
$C_{v}\times V^{-1}, U_{4}$ and their derivatives $\frac{\partial
  C_{v}}{\partial\beta}\times V^{-1}, \frac{\partial
  U_{4}}{\partial\beta}\times V^{-1}$ as a function of $\beta$ and
which extremum we have studied in each case.  They have been produced
by reweighting a run with $10^{6}$ iterations on a $12^{4}$ lattice at
$\beta=1.01024$. We recall that the critical coupling of U(1) lattice
gauge with Wilson action and periodic boundary conditions is
$\beta_{c}(\infty) \approx 1.011$.

\subsection{Derivatives of the free energy density}

\begin{figure}
\center
  \begin{tabular}{c}
  \epsfig{file=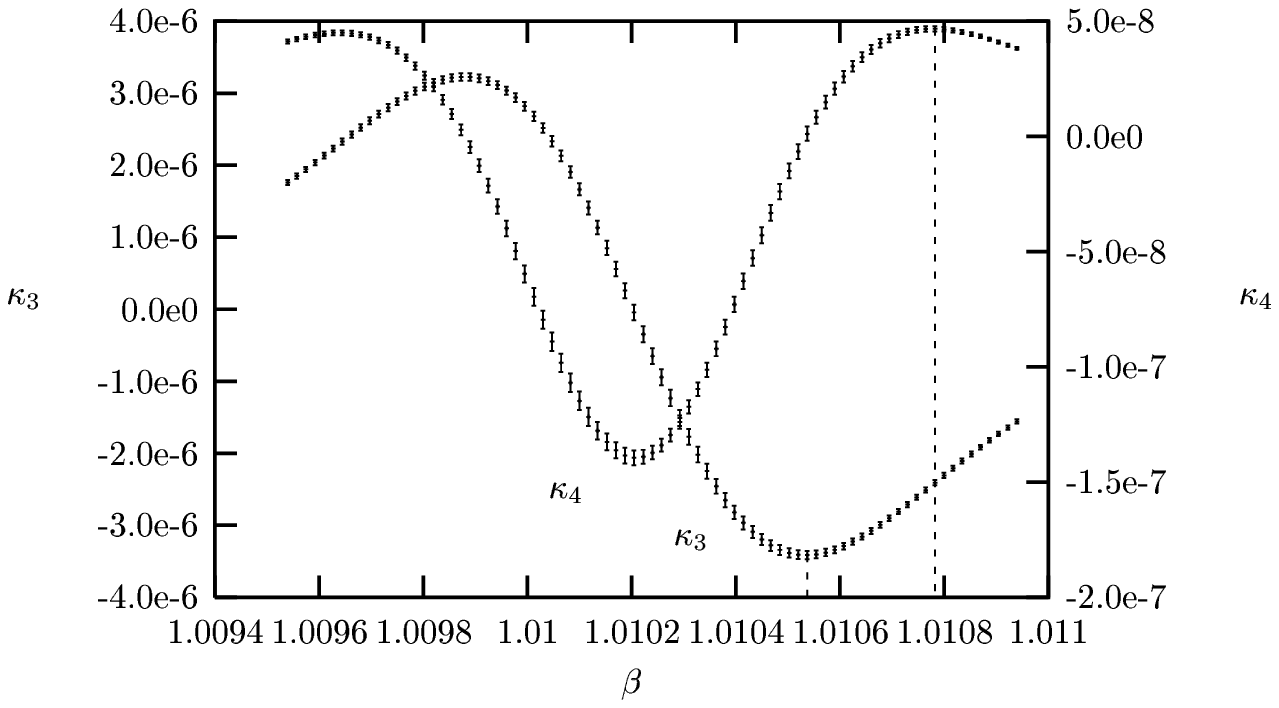, width=\linewidth} \\
  \epsfig{file=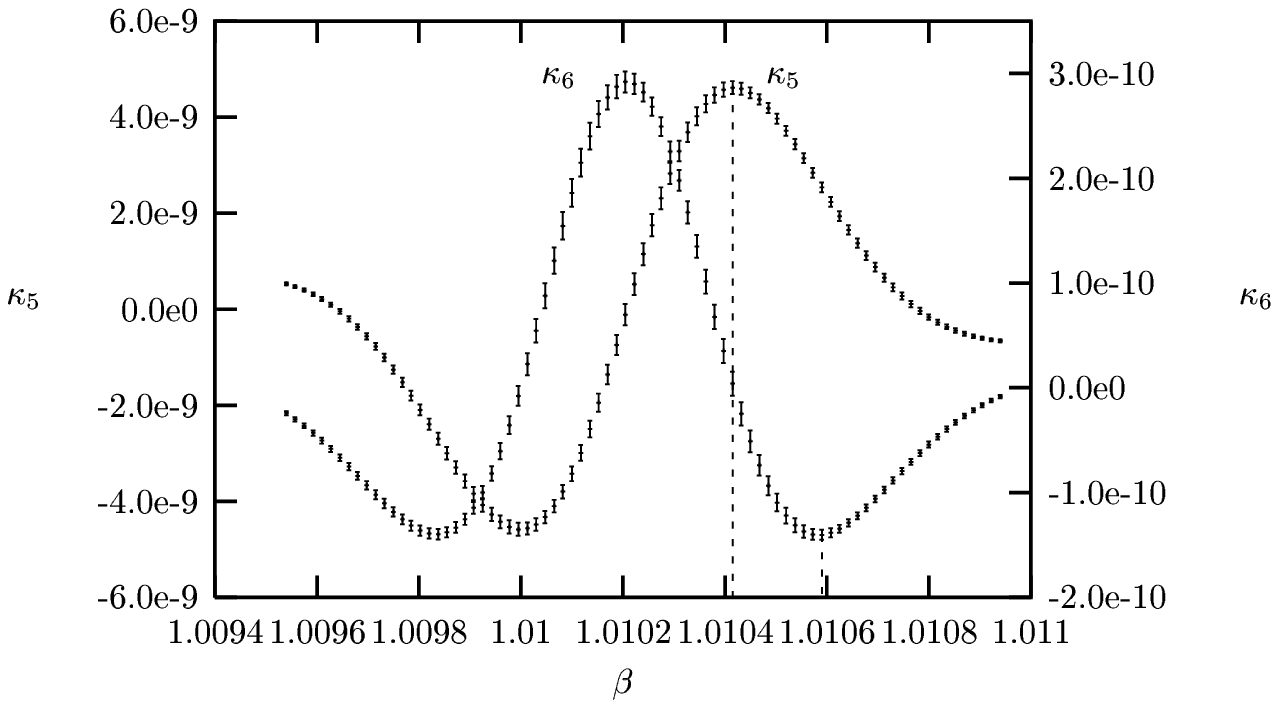, width=\linewidth}
  \end{tabular}
  \caption{The cumulants $\kappa_{n}\times V^{-n+1}$ on a $12^4$ lattice
           as a function of $\beta$.}
  \label{fig:kn}
\end{figure}

We have just seen that it is not always easy to interpret the scaling
corrections to the standard cumulants or their derivatives. The
interpretation would be more tractable if we could study cumulants
which are directly expressible in terms of the free energy. In
particular we can introduce the energy cumulants $\kappa_{n}(\beta,L)$
\cite{JK} which are defined through the Taylor expansion of the free
energy density $F(\beta,L)$:
\begin{eqnarray}
  F(\beta',L) & = & F(\beta,L) - \sum_{n=1}^{\infty}%
  \frac{(-1)^{n}}{n!}\kappa_{n}(\beta,L)(\beta'-\beta)^{n}
\end{eqnarray}
The finite-size scaling behavior of these energy cumulants is much simpler
since they are simply the derivatives of the free energy density:
\begin{eqnarray}
  &(-1)^{n+1}& \kappa_{n}(\beta,L) = %
  \frac{\partial^{n}F(\beta,L)}{\partial\beta^{n}} \hfill\nonumber\\
  &=& \frac{1}{\beta_{c}^{n}}L^{-d+\frac{n}{\nu}}\left(f_{0}^{(n)}(x)%
  + L^{-\omega}f_{1}^{(n)}(x)\right) + f_{ns}^{(n)}(\beta)
\end{eqnarray}
The analytic piece $f_{ns}^{(n)}(\beta)$ can be neglected as soon as
$n \geq 3$.  The first three cumulants coincide with the central
moments:
\begin{eqnarray}
  \kappa_{1} & = & \left<E\right> \\
  \kappa_{2} & = & V\left(\left<E^{2}\right>-\left<E\right>^{2}\right)%
  \quad = \quad  \mu_{2} \quad  = \quad  C_{v}/\beta^{2} \\
  \kappa_{3} & = & V^{2}\left<\left(E-\left<E\right>\right)^{3}\right>%
  \quad = \quad \mu_{3}
\end{eqnarray}
The higher-order cumulants can be expressed as non-linear combinations of the
central moments $\mu_{n} =
V^{n-1}\left<\left(E-\left<E\right>\right)^{n}\right>$. We shall introduce in
our analysis the cumulants $\kappa_{n}$ up to sixth order:
\begin{eqnarray}
   \kappa_{4} & = & \mu_{4} -3V\mu_{2}^{2} \\ 
   \kappa_{5} & = & \mu_{5} -10V\mu_{2}\mu_{3} \\ 
   \kappa_{6} & = & \mu_{6} -15V\mu_{2}\mu_{4}%
   -10V\mu_{3}^{2} +30V^{2}\mu_{2}^{3}
\end{eqnarray}
Fig.~\ref{fig:kn} displays the plots of the cumulants $\kappa_{n}\times
V^{-n+1}$ on a $12^4$ lattice as a function of $\beta$ and which extremum we
have studied in each case.

These energy cumulants have not been studied very often in finite-size
scaling analyses of phase transitions. In lower dimensions
and for spin models the emphasis is of course on magnetic cumulants
and the standard set of cumulants is natural in this context. The
energy cumulants $\kappa_{n}$ are more suitable to the analysis of
four-dimensional field theories.

\section{Simulation details}
\label{simul}

Previous studies \cite{KLA,ROI} provide us already with a good
knowledge of the locations of pseudo-critical couplings at eight
lattice sizes. We have therefore decided to increase the statistics at
several couplings inside the error bars of these pseudo-critical
points. For each coupling we did at least two runs with hot and cold
starts, different random generators with large periods, and statistics
of $5\cdot10^5$ configurations. For a complete overview of our
statistics we refer to Table \ref{tab:mcit}. The simulated
$\beta$-values are quite close to each other and we can use the
spectral density method \cite{FS} to study the pseudo-critical points
without noticeable extrapolation errors.

\begin{table}
\center
\begin{tabular}{|r|c|r|r|c|r|}
  L  & $\beta$  &  iterations         & L  & $\beta$ &  iterations       \\
\hline
  4  & 0.9785   &                     & 5  & 0.9940  &                   \\  
     &    $|$   & $5\times 10^{6}$    &    &   $|$   & $12\times 10^{6}$ \\
     & 0.9795   &                     &    & 0.9952  &                   \\ 
\hline 
  6  & 1.0014   &                     & 7  & 1.0052  &                   \\  
     &    $|$   & $7\times 10^{6}$    &    &   $|$   & $4\times 10^{6}$  \\
     & 1.0020   &                     &    & 1.0055  &                   \\  
\hline 
  8  & 1.0072   &                     & 9  & 1.0084  &                   \\  
     &   $|$    & $5.5\times 10^{6}$  &    &   $|$   & $4.5\times 10^{6}$\\
     & 1.0076   &                     &    & 1.0088  &                   \\
\hline 
  10 & 1.0092   & $1\times 10^{6}$    &12 & 1.0101   & $1\times 10^{6}$  \\  
     & 1.0093   & $1\times 10^{6}$    &   & 1.0102   & $1\times 10^{6}$  \\    
     & 1.0094   & $1.5\times 10^{6}$  &   & 1.0103   & $1\times 10^{6}$  \\  
     & 1.0095   & $1\times 10^{6}$    &\textsc{Mc} & 1.01024&$4.2\times 10^6$\\ 
     &          &                     &\textsc{Mc} & 1.0102& $1.4\times 10^6$\\
\end{tabular}
\caption{Number of configurations generated on different lattice
         sizes. \textsc{Mc} indicates multicanonical data.
         Simulations  were performed in steps of $10^{-4}$.}
\label{tab:mcit}
\end{table}

\begin{figure}
\center
   \epsfig{file=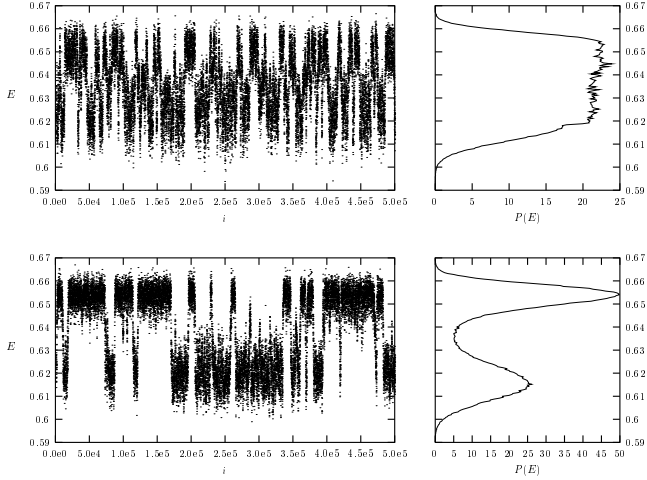, width=\columnwidth}
    \caption{The histories in multicanonical and heatbath simulations on
      a $12^4$ lattice and the corresponding histograms normalized to
      unit area.}
    \label{fig:mc}
\end{figure}

We have of course simulated the full U(1) group in double precision.
We used two different programs to generate our configurations. One is
an improved heatbath algorithm, the other one is capable of
multicanonical updates (\textsc{Mc}) combined with overrelaxation
\cite{BN,KLA}. Both programs were tuned to yield acceptance rates of
about 60\%. The multicanonical algorithm was used in addition to the
heatbath algorithm on $12^4$ lattices. To produce the histogram needed
for the multicanonical simulation we generated $1.5\cdot 10^5$
configurations using an overrelaxed Metropolis algorithm. For
$\beta=1.01020$ the multicanonically reweighted histograms of the runs
at $\beta=1.01024$ were used as the input.
In all \textsc{Mc} runs we used eight Metropolis updates
followed by one overrelaxation step per link. A drastic reduction of
the tunneling time is evident in figure \ref{fig:mc}.
On smaller lattice sizes the heatbath algorithm is sufficient to
generate statistics that allow a precise determination of all
observables. 

\begin{figure}
\center
  \epsfig{file=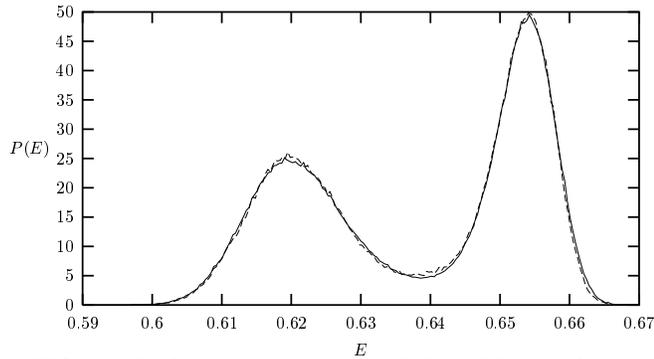, width=\linewidth}
    \caption{The histograms in \textsc{Mc} (solid) and heatbath
      simulations (dashed) on a $12^4$ lattice.}
    \label{fig:hgmvgl}
\end{figure}

For safety reasons all results on our data sets were cross-checked
with two independently developped evaluation programs. As a test of
consistency we reweighted the multicanonical histogram to a coupling
where we had heatbath data (see figure \ref{fig:hgmvgl}). Though we
did not apply any smoothing algorithms to the histograms no relevant
deviation can be observed.

To calculate the errors of our observables we proceeded as follows: We
first divided our runs into five bins and calculated the error in each
run by binning. The error for each lattice size was then calculated by
a $\chi^2$-- fit to a constant of all runs' binning results. In other
words we did not recombine the histograms at the pseudo-critical
points and used each run as an independent sample. Secondly we
performed a jacknife error analysis in the same manner but with ten
bins. The larger error was taken into account. The jacknife error
appeared to be usually larger, especially for the cumulants'
derivatives, in heatbath data at all lattice sizes except $L=12$. For
$L=12$ the binning error was very large in heatbath data while no
significant difference could be seen between the two methods in
\textsc{Mc} data. We interpret this phenomenon as a consequence of the
lack of tunnelling events within a bin on a $12^{4}$ lattice with a
local heatbath algorithm. We observe just a few phase flips in
$10^{5}$ iterations.  It is our experience that one needs at least a
total of $O(10^{2})$ flips in order to control the statistical errors
on the observables.

\section{Measurements}
\label{measure}

The measurements of the standard cumulants and their derivatives are displayed
in Table~\ref{tab:std} for all the lattice sizes that we have studied and the
results for the cumulants $\kappa_{n} (3\leq n \leq 6)$ are gathered in
Table~\ref{tab:kn}.

\begin{table}
\center
\begin{tabular}{|r|l|l|l|l|} 
  $L$ &  $\beta_{c}(C_v)$ &  $C_{V,max}\times V^{-1}$ &%
  $\beta_{c}(\partial C_v/\partial\beta)$ &%
  $V^{-1}\partial C_V/\partial\beta$ \\ 
\hline
  4 &  0.9791(2)  & .2205(5)E-02 &  0.9901(2)  & .0803(8) \\ 
  5 &  0.99466(5) & .1390(3)E-02 &  1.00035(6) & .101(1)  \\
  6 &  1.00172(4) & .955(3)E-03  &  1.00499(4) & .124(3)  \\
  7 &  1.00532(5) & .703(3)E-03  &  1.00735(5) & .150(7)  \\
  8 &  1.00744(4) & .551(4)E-03  &  1.00877(4) & .186(10) \\
  9 &  1.00862(3) & .453(3)E-03  &  1.00951(3) & .230(14) \\
 10 &  1.00939(5) & .382(4)E-03  &  1.01002(6) & .265(11) \\
 12 &  1.010232(7)& .302(2)E-03  &  1.010567(8)& .429(5)  \\ 
\hline
  $L$ &  $\beta_{c}(U_2)$ &  $U_{2,min}$ &%
  $\beta_{c}(\partial U_2/\partial\beta)$ &%
  $V^{-1}\partial U_2/\partial\beta$ \\ 
\hline
  4 &  0.9764(2)  & -.670(2)E-02  &  0.9882(2)   & -.1574(15)E-03\\
  5 &  0.99371(5) & -.3857(9)E-02 &  0.99961(6)  & -.752(8)E-04  \\
  6 &  1.00132(4) & -.2525(7)E-02 &  1.00466(4)  & -.423(9)E-04  \\
  7 &  1.00512(5) & -.1802(8)E-02 &  1.00719(5)  & -.269(12)E-04 \\
  8 &  1.00734(4) & -.1387(9)E-02 &  1.00868(4)  & -.191(10)E-04 \\
  9 &  1.00855(3) & -.1122(8)E-02 &  1.00945(3)  & -.146(9)E-04  \\
 10 &  1.00935(5) & -.944(10)E-03 &  1.00998(6)  & -.110(6)E-04  \\
 12 &  1.010217(6)& -.734(4)E-03  &  1.010553(7) & -.841(9)E-05  \\ 
\hline
 $L$ &  $\beta_{c}(U_4)$ &  $U_{4,min}$ &%
 $\beta_{c}(\partial U_4/\partial\beta)$ &%
 $V^{-1}\partial U_4/\partial\beta$ \\ 
\hline
  4 &  0.9753(2)  & -.897(2)E-02   &  0.9870(2)   & -.211(2)E-03   \\
  5 &  0.99327(5) & -.5158(11)E-02 &  0.99918(6)  & -.1006(10)E-03 \\
  6 &  1.00111(4) & -.3372(10)E-02 &  1.00446(4)  & -.565(12)E-04  \\
  7 &  1.00501(5) & -.2404(11)E-02 &  1.00708(5)  & -.359(15)E-04  \\
  8 &  1.00728(4) & -.1851(12)E-02 &  1.00861(4)  & -.255(13)E-04  \\
  9 &  1.00851(3) & -.1513(11)E-02 &  1.00941(3)  & -.195(12)E-04  \\
 10 &  1.00932(5) & -.1254(14)E-02 &  1.00996(4)  & -.148(9)E-04   \\
 12 &  1.010203(7)& -.979(6)E-03   &  1.010538(9) & -.1122(12)E-04 \\ 
\end{tabular}
\caption{Pseudo-critical couplings and extrema of $C_{v}, U_{2}, U_{4}$ and
          their derivatives as a function of the lattice size $L$.}
\label{tab:std}
\end{table}

\begin{table}
\center
\begin{tabular}{|r|l|l|l|l|} 
  $L$ &$\beta_{c}(\kappa_3)$ & $\kappa_3\times V^{-2}$ &%
  $\beta_{c}(\kappa_4)$ & $\kappa_4\times V^{-3}$\\ 
\hline
   4  &  0.9902(2)  &  -0.511(6)E-04  &  0.9983(2)   & 0.156(3)E-05\\
   5  &  1.00035(6) &  -0.264(3)E-04  &  1.00452(6)  & 0.659(9)E-06\\
   6  &  1.00499(4) &  -0.156(4)E-04  &  1.00737(5)  & 0.334(9)E-06\\
   7  &  1.00735(5) &  -0.102(5)E-04  &  1.00882(5)  & 0.191(9)E-06\\
   8  &  1.00877(4) &  -0.74(4)E-05   &  1.00973(4)  & 0.126(7)E-06\\
   9  &  1.00951(3) &  -0.57(4)E-05   &  1.01016(3)  & 0.90(6)E-07\\
  10  &  1.01002(6) &  -0.43(2)E-05   &  1.01048(6)  & 0.68(8)E-07\\
  12  &  1.010564(9)&  -0.338(4)E-05  &  1.010807(10)& 0.460(7)E-07\\
\hline
  $L$ &$\beta_{c}(\kappa_5)$ & $\kappa_5\times V^{-4}$ &%
  $\beta_{c}(\kappa_6)$ & $\kappa_6\times V^{-5}$\\
\hline
  4   &  0.9851(2)  & 0.385(7)E-06 &  0.9907(2)  &  -0.278(6)E-07\\
  5   &  0.99827(6) & 0.131(3)E-06 &  1.00110(6) &  -0.770(13)E-08\\
  6   &  1.00381(4) & 0.55(2)E-07  &  1.00543(4) &  -0.273(8)E-08\\
  7   &  1.00663(5) & 0.27(2)E-07  &  1.00764(5) &  -0.117(6)E-08\\
  8   &  1.00830(4) & 0.162(13)E-07&  1.00895(4) &  -0.62(4)E-09\\
  9   &  1.00919(3) & 0.11(9)E-07  &  1.00964(3) &  -0.38(3)E-09\\
  10  &  1.00979(6) & 0.71(8)E-08  &  1.01011(6) &  -0.24(4)E-09\\
  12  &  1.010446(8)& 0.452(9)E-08 &  1.010611(9)&  -0.137(3)E-09\\
\end{tabular}
\caption{Pseudo-critical couplings and extrema of $\kappa_{n}$ cumulants as a
          function of the lattice size $L$.}
\label{tab:kn}
\end{table}

It can be read from the tables that the statistical accuracy of the
measurements of the pseudo-critical couplings and of the cumulants
extrema has been improved by roughly one-order of magnitude with
respect to previous studies. In particular the relative accuracy of
our results for the pseudo-critical couplings on the $12^{4}$ lattice
is now below $10^{-5}$, which is nearly comparable to the best results
of numerical studies of lower-dimensional spin models.

In the following sections we shall present a finite-size scaling
analysis of these results. But our analysis of corrections to scaling
will be conventional and far from exhaustive. With our measurements
given in raw form, without any interpretation, the reader has the
possibility to do alternative analyses.

\section{Pseudo-critical couplings}
\label{coupling}

The pseudo-critical couplings are expected to follow the asymptotic
finite-size scaling formula:
\begin{eqnarray}
  \label{fss}
  \beta_{c}(L)=\beta_{c}(\infty) + a L^{-\frac{1}{\nu}}
\end{eqnarray}
With data at eight lattice sizes it is possible to fit the three unknown
parameters for each cumulant independently. With our statistics it is even
possible to test the stability of the fits with respect to the lattice size.
Table~\ref{tab:gc} displays the results of such fits for the standard
cumulants and their derivatives.

\begin{table}
\center
\begin{tabular}{|c|c|c|c|c|c|} 
$Cumulant$ & $L$ & $\chi^2$ & $\nu^{-1}$ & a & $\beta_{c}(\infty)$ \\  
\hline                          
 $C_{v}$%
      & $L \geq$  4 &   2.42     &   2.999(17) &  -2.091(56) &   1.01145(3) \\
      & $L \geq$  5 &   1.42     &   3.037(23) &  -2.226(81) &   1.01141(3) \\
      & $L \geq$  6 &  0.731     &   3.111(46) &  -2.54(21)  &   1.01135(5) \\
      & $L \geq$  7 &  0.379     &   3.23(11)  &  -3.17(65)  &   1.01128(7) \\
\hline                          
 $U_{2}$%
      & $L \geq$  4 &   1.29     &   3.090(16) &  -2.554(64) &   1.01140(2) \\
      & $L \geq$  5 &   1.04     &   3.112(22) &  -2.648(92) &   1.01138(3) \\
      & $L \geq$  6 &  0.829     &   3.161(44) &  -2.89(23)  &   1.01134(4) \\
      & $L \geq$  7 &  0.704     &   3.26(11)  &  -3.48(68)  &   1.01128(7) \\
\hline                          
 $U_{4}$%
      & $L \geq$  4 &   1.24     &   3.116(16) &  -2.727(67) &   1.01139(2) \\
      & $L \geq$  5 &   1.08     &   3.136(21) &  -2.818(97) &   1.01137(3) \\
      & $L \geq$  6 &  0.878     &   3.184(43) &  -3.07(24)  &   1.01133(4) \\
      & $L \geq$  7 &  0.816     &   3.28(11)  &  -3.66(71)  &   1.01127(7) \\
\hline                          
 $\frac{\partial C_{v}}{\partial\beta}$%
     & $L \geq$  4 &   1.43     &   3.012(26) &  -1.398(58) &   1.01136(3) \\
     & $L \geq$  5 &   1.01     &   3.057(37) &  -1.509(91) &   1.01133(3) \\
     & $L \geq$  6 &  0.715     &   3.140(71) &  -1.75(22)  &   1.01128(5) \\
     & $L \geq$  7 &  0.730     &   3.26(17)  &  -2.22(72)  &   1.01123(7) \\
\hline                          
 $\frac{\partial U_{2}}{\partial\beta}$%
     & $L \geq$  4 &   1.12     &   3.096(24) &  -1.705(66) &   1.01133(2) \\
     & $L \geq$  5 &  0.836     &   3.133(35) &  -1.81(11)  &   1.01131(3) \\
     & $L \geq$  6 &  0.704     &   3.196(67) &  -2.03(24)  &   1.01128(4) \\
     & $L \geq$  7 &  0.898     &   3.28(16)  &  -2.37(73)  &   1.01124(7) \\
\hline                          
 $\frac{\partial U_{4}}{\partial\beta}$%
     & $L \geq$  4 &  0.724     &   3.142(24) &  -1.904(71) &   1.01131(3) \\
     & $L \geq$  5 &  0.687     &   3.165(34) &  -1.98(11)  &   1.01130(3) \\
     & $L \geq$  6 &  0.613     &   3.218(66) &  -2.18(26)  &   1.01127(4) \\
     & $L \geq$  7 &  0.732     &   3.30(16)  &  -2.57(77)  &   1.01123(7) \\
\end{tabular}
\caption{Independent second-order finite-size scaling fits to the
         pseudo-critical couplings of standard cumulants and
         their derivatives.}
\label{tab:gc}
\end{table}

\begin{figure}
\center
  \epsfig{file=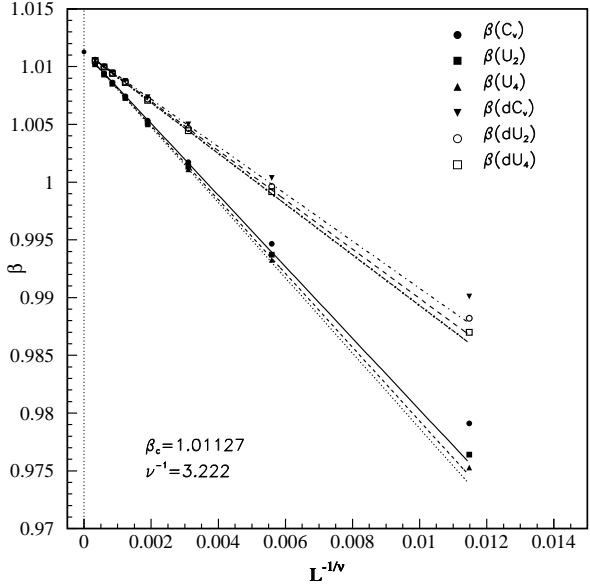, width=\linewidth}
    \caption{Combined fit to the $L\geq 6$ data points of the pseudo-critical
             couplings of the standard cumulants and their derivatives.}
    \label{fig:allgc}
\end{figure}

The fits are clearly not stable. The statistical errors are small
enough to show the systematic decrease of the critical exponent $\nu$
with the lattice size $L$. There is also a slight systematic decrease
of the infinite-volume limit of the critical coupling
$\beta_{c}(\infty)$. Moreover the fits are pretty consistent for all
cumulants. Therefore we can try a combined fit to the pseudo-critical
couplings of all standard cumulants and their derivatives, with a
common value of $\nu$ and $\beta_{c}(\infty)$. Such a fit is
meaningful since these cumulants are defined in terms of algebraically
independent combinations of the moments of the plaquette energy
distribution. We can also test the stability of this combined fit with
respect to the lattice size. The parameters of these fits are given in
Table~\ref{tab:allgc}.  Fig.~\ref{fig:allgc} displays the result of
the combined fit to the $L\geq 6$ data points. Errors on the data
points are much smaller than the marker sizes.

\begin{table}
\center
\begin{tabular}{|c|c|c|c|} 
$L$ & $\chi^2$ & $\beta_{c}(\infty)$ & $\nu^{-1}$  \\  
\hline
 $L \geq$  4 &   5.86     &   1.011348(8)  &   3.1059(72) \\
 $L \geq$  5 &   4.09     &   1.011317(10) &   3.144(10)  \\
 $L \geq$  6 &   1.81     &   1.011269(14) &   3.222(19)  \\
 $L \geq$  7 &  0.692     &   1.011224(19) &   3.329(35)  \\
 $L \geq$  8 &  0.799     &   1.011235(29) &   3.304(62)  \\
 $L \geq$  9 &  0.872E-01 &   1.011206(43) &   3.4249(69) \\
\end{tabular}
\caption{Combined asymptotic finite-size scaling fits to the pseudo-critical
         couplings of the cumulants $C_{v}$, $U_{2}$, $U_{4}$,
         $\frac{\partial C_{v}}{\partial\beta}$,
         $\frac{\partial U_{2}}{\partial\beta}$ and
         $\frac{\partial U_{4}}{\partial\beta}$. }
\label{tab:allgc}
\end{table}

\begin{figure}
\center
  \epsfig{file=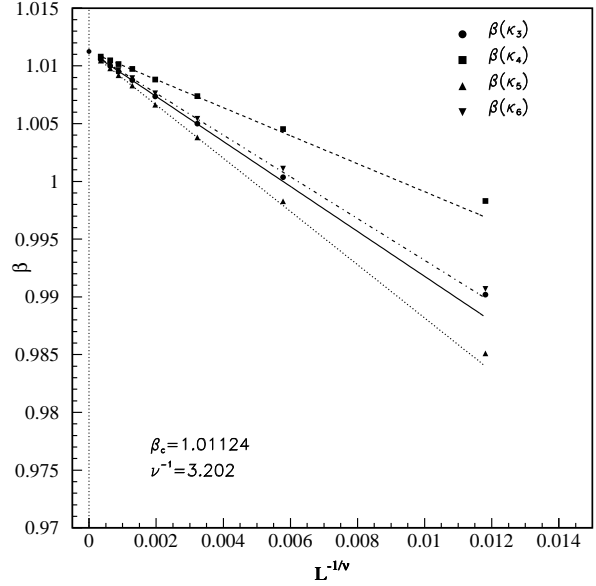, width=\linewidth}
    \caption{Combined fit to the $L\geq 6$ data points of the pseudo-critical
             couplings of the cumulants $\kappa_{n}$.}
    \label{fig:allkn}
\end{figure}

We can repeat the same kind of analysis for the energy cumulants
$\kappa_{n}$. The independent fits are shown in Table~\ref{tab:bkn}
and the combined fits in Table~\ref{tab:allkn}. The result of the
combined fit to the $L\geq 6$ data points is also displayed in
Fig.~\ref{fig:allkn}.

\begin{table}
\center
\begin{tabular}{|c|c|c|c|c|c|} 
$Cumulant$ & $L$ & $\chi^2$ & $\nu^{-1}$ & a & $\beta_{c}(\infty)$ \\  
\hline                          
 $\kappa_{3}$%
      & $L \geq$  4 &   1.72     &   3.005(27) &  -1.381(58) &   1.01136(3) \\
      & $L \geq$  5 &   1.05     &   3.060(38) &  -1.515(92) &   1.01132(3) \\
      & $L \geq$  6 &  0.720     &   3.146(72) &  -1.77(23)  &   1.01128(5) \\
      & $L \geq$  7 &  0.711     &   3.28(17)  &  -2.26(74)  &   1.01122(7) \\
\hline                          
 $\kappa_{4}$%
      & $L \geq$  4 &   1.18     &   3.011(44) &  -0.858(60) &   1.01129(3) \\
      & $L \geq$  5 &  0.872     &   3.078(62) &  -0.959(97) &   1.01127(3) \\
      & $L \geq$  6 &  0.724     &   3.20(13)  &  -1.19(27)  &   1.01123(5) \\
      & $L \geq$  7 &  0.812     &   3.38(28)  &  -1.69(93)  &   1.01119(7) \\
\hline                          
 $\kappa_{5}$%
      & $L \geq$  4 &   1.02     &   3.083(22) &  -1.876(65) &   1.01133(3) \\
      & $L \geq$  5 &   1.08     &   3.063(31) &  -1.815(92) &   1.01134(3) \\
      & $L \geq$  6 &  0.684     &   3.139(60) &  -2.08(22)  &   1.01130(5) \\
      & $L \geq$  7 &  0.721     &   3.24(15)  &  -2.51(68)  &   1.01125(8) \\
\hline                          
 $\kappa_{6}$%
      & $L \geq$  4 &  0.843     &   3.103(28) &  -1.512(67) &   1.01129(3) \\
      & $L \geq$  5 &  0.887     &   3.079(40) &  -1.454(95) &   1.01130(3) \\
      & $L \geq$  6 &  0.485     &   3.174(77) &  -1.72(24)  &   1.01126(5) \\
      & $L \geq$  7 &  0.489     &   3.29(19)  &  -2.137(64) &   1.01122(7) \\
\end{tabular}
\caption{Independent second-order finite-size scaling fits to the 
         pseudo-critical couplings of the $\kappa_{n}$ cumulants.}
\label{tab:bkn}
\end{table}

\begin{table}
\center
\begin{tabular}{|c|c|c|c|} 
$L$ & $\chi^2$ & $\beta_{c}(\infty)$ & $\nu^{-1}$ \\  
\hline
 $L \geq$  4 &   2.41     &   1.011299(12)  &   3.084(14) \\
 $L \geq$  5 &   1.59     &   1.011288(13)  &   3.105(18) \\
 $L \geq$  6 &  0.718     &   1.011244(17)  &   3.202(31) \\
 $L \geq$  7 &  0.442     &   1.011206(25)  &   3.319(63) \\
 $L \geq$  8 &  0.564     &   1.011220(37)  &   3.27(11)  \\
 $L \geq$  9 &  0.856E-02 &   1.011191(53)  &   3.429(14) \\
\end{tabular}
\caption{Combined asymptotic finite-size scaling fits to the pseudo-critical 
         couplings of the cumulants $\kappa_{3}$, $\kappa_{4}$, 
         $\kappa_{5}$ and $\kappa_{6}$.}
\label{tab:allkn}
\end{table}

The critical values of $\nu$ and $\beta_{c}(\infty)$ extracted from the
cumulants $\kappa_{n}$ are very similar to those extracted from the standard
cumulants or their derivatives. Studying the pseudo-critical couplings of the
cumulants $\kappa_{n}$ does not bring any significant improvement over the
standard cumulants. The pseudo-critical couplings are not very sensitive to
the corrections to asymptotic scaling. Even if the finite-size scaling
violations are clearly visible on Figs.~\ref{fig:allgc} and \ref{fig:allkn} at
the small lattice sizes $L=4$ or $5$, it would not be possible to introduce
additional parameters in the fits.

\section{Cumulants' extrema}
\label{cumul}

In this section we will present some selected analyses of the finite
size scaling behaviour of the cumulants we calculated.  As the main
result we can state that the finite-size scaling behavior of all
cumulants is consistent with a first-order transition. However we
observe a difference in the corrections to scaling of the specific
heat and the Binder cumulant on one hand and their derivatives and the
energy cumulants on the other hand. Up to the lattice sizes we
calculated the first do not follow yet the expected pattern in a first
order phase transition while the other show a clear first order
behaviour. To demonstrate this we will present selected fits to our
data. In all pictures in the following subsections the solid curves
denote the fit to the ansatz which yields the best results.

\subsection{Asymptotic scaling}

It is certainly not possible to describe the data without taking into account
the corrections to scaling. However it is instructive to define, for any
cumulant $\kappa$, an ''effective critical exponent'' $\nu_{\kappa}(L)$ by
fitting the data at scales $L, L+1$ and $L+2$ to the asymptotic form of the
scaling ans\"atze of sect.~\ref{frame}:
\begin{eqnarray}
  \label{fss2a}
FSS2a:\qquad\qquad  \kappa(L) & = & b L^{\tau}
\end{eqnarray}
We shall refer to this ansatz as $FSS2a$. The relation between the critical
exponents $\tau$ and $\nu$ depends of course upon the cumulant. The results
are displayed in Table~\ref{tab:as2}.

\begin{table}
\center
\begin{tabular}{|c|c|c|c|c|}
$L$ & $C_{v}$     & $U_{4}$      & $\frac{\partial C_{v}}{\partial\beta}$  &%
    $\frac{\partial U_{4}}{\partial\beta}$ \\
\hline
9 & 3.299(17)   &   3.250(17)  &  3.467(56)  &   3.402(63) \\
8 & 3.177(28)   &   3.132(27)  &  3.19(11)   &   3.19(12)  \\
7 & 3.121(16)   &   3.070(17)  &  3.23(11)   &   3.183(97) \\
6 & 3.030(12)   &   2.942(12)  &  3.123(64)  &   3.064(59) \\
5 & 2.983(7)    &   2.856(7)   &  3.050(36)  &   2.962(32) \\
4 & 2.968(5)    &   2.788(5)   &  3.016(18)  &   2.906(16) \\
\hline
$L$ & $\kappa_{3}$ & $\kappa_{4}$ &$\kappa_{5}$ & $\kappa_{6}$ \\
\hline
9 & 3.479(63)  &   3.422(57)  &  3.405(55) &  3.416(47) \\
8 & 3.19(11)   &   3.30(13)   &  3.27(12)  &  3.30(11)  \\
7 & 3.22(12)   &   3.247(80)  &  3.282(90) &  3.246(62) \\
6 & 3.125(66)  &   3.139(51)  &  3.135(59) &  3.126(39) \\
5 & 3.048(38)  &   3.075(30)  &  3.055(36) &  3.061(22) \\
4 & 3.019(20)  &   3.046(20)  &  3.038(18) &  3.045(15) \\
\end{tabular}
\caption{Critical exponent $\nu^{-1}$ extracted from lattice sizes
         $L$, $L+1$ and $L+2$ for each cumulant extremum with
         the ansatz $FSS2a$.}
\label{tab:as2}
\end{table}

We observe again in Table~\ref{tab:as2} a systematic decrease of the
critical exponent $\nu$ with the lattice size $L$. The effective
critical exponent $\nu(L)$ is pretty consistent at each scale $L$
across the cumulants $\kappa_{n}$ and with the values found in the
combined fits of the pseudo-critical couplings. There is however some
dispersion among the standard cumulants and their derivatives not
observed in the analysis of pseudo-critical couplings. In fact the
$\chi^2$ of these asymptotic fits is quite high even with 2
parameters and only 3 data points. We see here a first manifestation
of the expected improved scaling behavior of the energy cumulants
$\kappa_{n}$.

\subsection{Corrections to scaling}

The asymptotic scaling fits for the cumulants' extrema as well as for the
pseudo-critical couplings hint at a first-order transition. Our working
hypothesis will be to describe the data with scaling corrections to a
first-order transition. These corrections are expected to be expressible as a
series expansion in the inverse volume $V^{-1}$. Since we limit ourselves to
three-parameter fits, we shall introduce the two fitting ans\"atze:
\begin{eqnarray}
  \label{ffs1a}
  FSS1a:\qquad\qquad  \kappa(L)\times V^{-k} & = & a + b L^{-d}  \\
  \label{ffs1b}
  FSS1b:\qquad\qquad  \kappa(L)\times V^{-k} & = & a + b L^{-d} + c L^{-2d}  
\end{eqnarray}
The volume normalization factor is chosen such that the fitting ansatz
makes sense. With the definitions of sect.~\ref{frame} we have $k=0$
for $U_{2}$ and $U_{4}$, $k=1$ for $C_{v}, \frac{\partial
  U_{2}}{\partial\beta}, \frac{\partial U_{4}}{\partial\beta}$, $k=2$
for $\frac{\partial C_{v}}{\partial\beta}$ and $k=n-1$ for
$\kappa_{n}$.

We shall need another, more phenomenological fitting ansatz:
\begin{eqnarray}
  \label{fss2b}
  FSS2b:\qquad\qquad  \kappa(L)\times V^{-k} & = & a + b L^{-\omega} 
\end{eqnarray}
The volume normalization factor $V^{-k}$ is by definition the same as
for the first-order fitting ans\"atze. Therefore the exponent $\omega$
parameterizes the corrections to an asymptotic first-order behaviour.
Introducing such an effective exponent allows a more flexible
description of the corrections. One should find that $\omega \rightarrow
d$ when we reach the asymptotic regime.

\begin{figure}
\center
  \epsfig{file=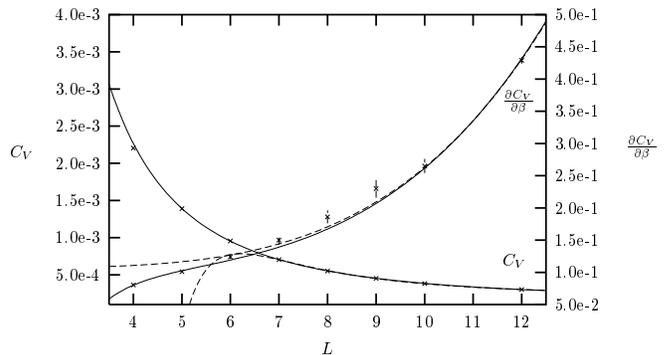, width=\linewidth}
  \caption{Fits to the specific heat and its derivative.}
  \label{fig:cvfit}
\end{figure}

The best $FSS1b$ fit to the specific heat is represented by the dashed
line in Fig.~\ref{fig:cvfit}. Obviously it cannot describe the
smallest lattice sizes.  On the other hand the fitting ansatz $FSS2b$
describes our data on the specific heat perfectly.
Our data on $\partial C_{v} / \partial \beta$ have the largest errors of
all the results we produced. Thus the quality of the best fits with
$FSS1b$ (solid line) or $FSS1a$ (dashed line) is not good.

\begin{figure}
\center
  \epsfig{file=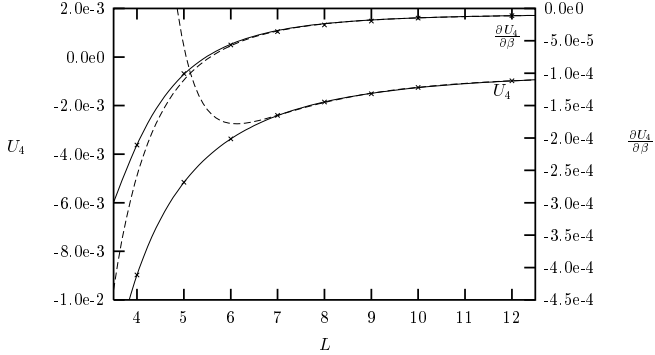, width=\linewidth}
  \caption{Fits to the Binder Cumulant and its derivative.}
  \label{fig:u4fit}
\end{figure}

The ansatz $FSS2b$ is also able to reproduce the data on the Binder
cumulant quite accurately (solid line in Fig.~\ref{fig:u4fit}). We
observe again that ansatz $FSS1b$ (dashed line) fails to reproduce the
smallest lattice sizes. The description of the derivative $\partial
U_{4} / \partial \beta$ by $FSS1b$ (solid line) is much better and
$FSS1a$ gives already a good fit for all but the smallest
lattice sizes (dashed line).
The results for $U_{2}$ and its derivative are quite similar to
$U_{4}$ and its derivative and will not be reproduced here.

\begin{figure}
\center
  \epsfig{file=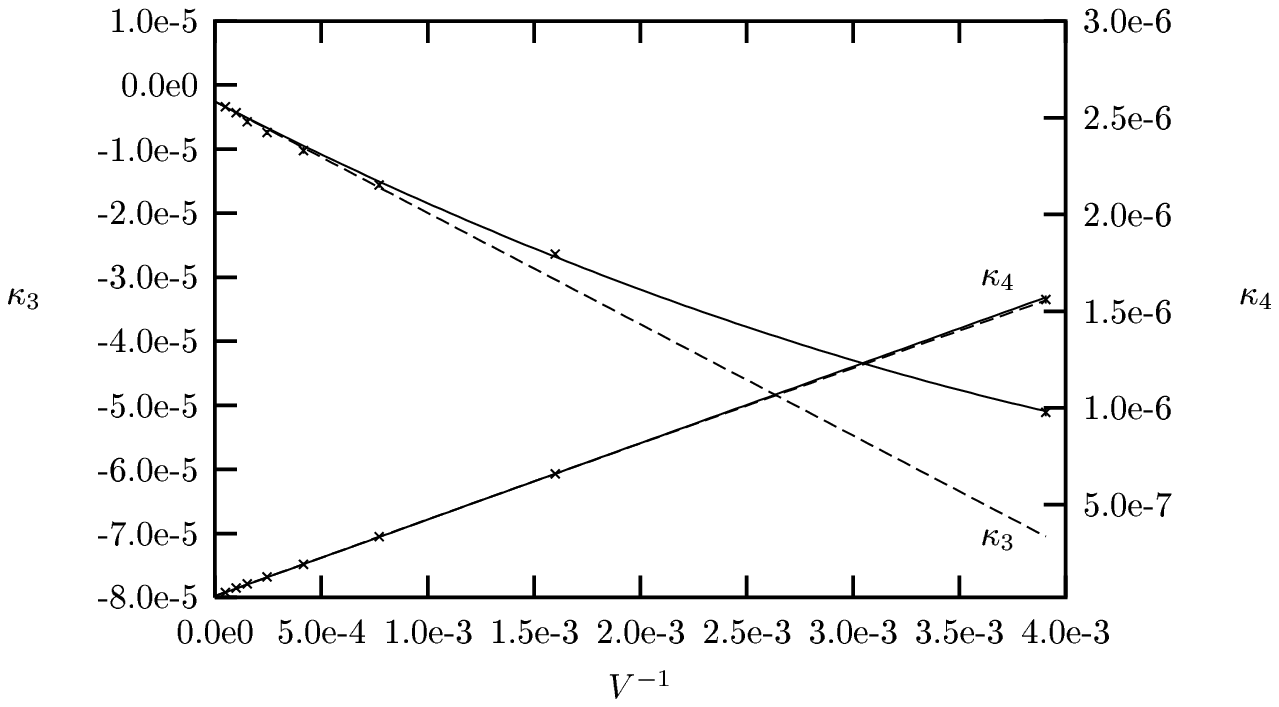, width=\linewidth}
  \epsfig{file=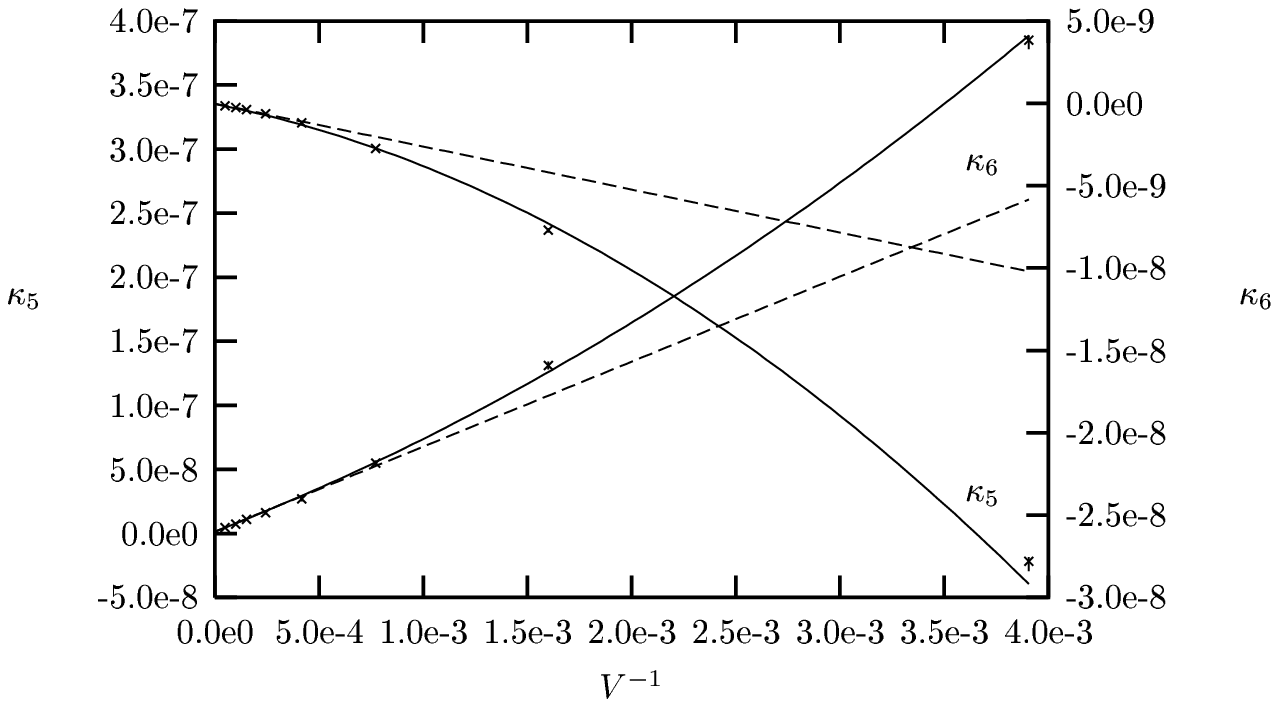, width=\linewidth}
  \caption{Fits to the energy cumulants $\kappa_{n}$.}
  \label{fig:knfit}
\end{figure}

In the same way we find that the first-order ans\"atze can fit the
data on the energy cumulants $\kappa_{n}$. In Fig.~\ref{fig:knfit} all
solid lines represent the best fits with ansatz $FSS1b$ whereas all
dashed lines represent the best fits with ansatz $FSS1a$. The cumulant
$\kappa_{4}$ is a noticeable exception.  The asymptotic first-order
ansatz $FSS1a$ is already able to describe all data points and is
nearly indistinguishable from ansatz $FSS1b$. We have another, and more
vivid manifestation of the improved scaling behaviour of the energy
cumulants $\kappa_{n}$.

Table~\ref{tab:fss} gives a quantitative content to the previous
qualitative observations. This table contains the parameters $a$ and
$b$ of the best fits displayed in the figures. We always choose
the smallest lattice size with a $\chi^{2}$ per degree of
freedom less than 2.

\begin{table}
\center
\begin{tabular}{|c|c|c|c|c|c|} 
$Cum.$ & $Fit$ & $L$ & $\chi^2$ & a & b \\ 
\hline
$C_{v}$      & $FSS2b$  & $L \geq$  5 &  0.52     & .147(6)E-03  & .580(23)E-01    \\
$\frac{\partial C_{v}}{\partial\beta}$%
             & $FSS1b$  & $L \geq$  5 &  0.58     &  .149(4)E-04 & .121(5)       \\
$U_{4}$      & $FSS2b$  & $L \geq$  4 &  0.85     & -.545(11)E-03  & -.352(7)      \\
$\frac{\partial U_{4}}{\partial\beta}$%
             & $FSS1b$  & $L \geq$  4 &  1.79      & -.823(14)E-05  & -.632(11)E-01   \\
$\kappa_{3}$ & $FSS1b$  & $L \geq$  5 &  0.50     & -.244(6)E-05 & -.196(9)E-01  \\
$\kappa_{4}$ & $FSS1a$  & $L \geq$  4 &  0.94E-01 & -.270(8)E-07 & -.395(5)E-03  \\
$\kappa_{5}$ & $FSS1b$  & $L \geq$  4 &   1.26     & .143(15)E-08   & .631(22)E-04    \\
$\kappa_{6}$ & $FSS1b$  & $L \geq$  5 &  0.43     & -.334(71)E-10  & -.206(13)E-05   \\
\end{tabular}
\caption{Best fits to the cumulants' extrema.}
\label{tab:fss}
\end{table}

The main feature is the smallness of the ratios $\frac{a}{b}$. In the
first-order fits, the leading contribution $a$ is generally smaller
than the correction term $bL^{-d}$ up to the lattice size $L=12$.  It
is no surprise that the first-order nature of the transition is so
difficult to observe.

We can get order-of-magnitude estimates of the parameters of a
first-order transition by comparing the values of $a$ with the
predictions of the (very crude) double-gaussian approximation
\cite{CLB}. The height of the maximum of the specific heat should
increase linearly with $L^{d}$ as:
\begin{eqnarray}
  \label{cvfss1}
  C_{v,max} & = & V\frac{3\beta_{c}^{2}}{2}(e_{o}-e_{d})^{2} + O(1)
\end{eqnarray}
where $\beta_{c}$ is the infinite-volume critical coupling and
$e_{o}-e_{d}$ is the latent heat. We find 
\begin{eqnarray}
  \label{gap1}
  e_{o}-e_{d} \approx 0.029
\end{eqnarray}
The minimum of the Binder cumulant is predicted to be \cite{B90,LK}:
\begin{eqnarray}
  \label{u4fss1}
  U_{4,min}=-\frac{(e_{o}^{2}-e_{d}^{2})^{2}}{12(e_{o}e_{d})^{2}} + O(V^{-1})
\end{eqnarray}
We get another estimate of the latent heat which agrees pretty well with
(\ref{gap1}):
\begin{eqnarray}
  \label{gap2}
  e_{o}-e_{d} \approx 0.026
\end{eqnarray}

\section{Conclusion}

Our study shows that finite-size scaling violations are indeed present
in each observable for lattice sizes $4\leq L\leq 12$.  The observed
scaling violations are consistent for all observables: the critical
exponent $\nu$ systematically {\bf decreases} with the lattice size
$L$.  Asymptotic finite-size scaling of all cumulants and of their
pseudo-critical couplings yields consistently $\nu \approx 0.29$ for
$L\geq 9$ which points towards a first-order transition.

The scaling violations in the pseudo-critical couplings and the
cumulant values decrease slowly with the lattice size. This slow
variation, which is hard to unravel, can explain the claims for a
second-order transition with $\nu \approx 0.33$.  However the scaling
violations seem to decrease more rapidly for the derivatives
of the standard cumulants and for the cumulants $\kappa_{n}$. The
finite-size size behavior of all these cumulants can be completely
described in the range $4\leq L \leq 12$ by volume correction terms of
order $V^{-1}$ and $V^{-2}$.  The cumulant $\kappa_{4}$ is even
completely described by the asymptotic first-order formula.

There seems to be a correlation between the amount of scaling
violations in each cumulant extremum and the location of its
pseudo-critical-coupling. The closer the pseudo-critical coupling is,
at fixed lattice size, from the infinite-volume critical point, the
better the description of the extremum by a first-order transition.
All the cumulants we have studied follow this rule. Then we can make
two observations.

On the one hand, if we had chosen to study the extremum in each
cumulant which is farthest from $\beta_{c}(\infty)$ we would
certainly have found a behavior inconsistent with a first-order
transition and concluded to a second-order transition with a critical
exponent $\nu$ {\it greater} than 0.33. We suggest that this is the
origin of many claims for a large critical exponent $\nu$. Most of
them are not based on finite-size scaling analyses of the standard
cumulants, which all should give $\nu \leq 0.33$, but come from
various analyses done at couplings $\beta$ smaller than
$\beta_{C_{v}}(L)$, the pseudo-critical coupling of the specific heat
in the corresponding lattices.

On the other hand we could think of studying still higher-order
$\kappa_{n}$ cumulants to get closer to $\beta_{c}(\infty)$.  However
we would face two difficulties. First the statistical noise increases
with $n$. Secondly the extrapolation from the couplings where we
generated our configurations gets large. The solution might be to
generate the configurations for all lattice sizes at the
infinite-volume critical point and use the method developed
\cite{BK,BKM} for the study of first-order transitions in Potts
models.
  
However it is not clear how the proofs can be extended to the case of
the U(1) phase transition. Potts models have a discrete symmetry and a
local order parameter whereas the U(1) lattice gauge theory has a
continuous local symmetry and a non-local order parameter since the
proof of existence of the phase transition uses the Wilson criterion
\cite{FS}. In Potts models the physical correlation lengths in both
pure phases stay finite in the infinite volume limit whereas there
exist massless photons in the ordered phase of U(1). Clearly the U(1)
lattice gauge theory deserves as much numerical study as the
lower-dimensional spin models and theoretical understanding of the
scaling violations will be required before a definitive conclusion on
the nature of its phase transition.

\acknowledgments

One of us (C.R.) is grateful to the Freie Universit\"at Berlin for its
hospitality during a visit when this manuscript was completed and
acknowledges a financial support by the Graduiertenkolleg
``Strukturuntersuchungen, Pr\"azisionstests und Erweiterungen des
Standardmodells der Elementarteilchenphysik''.

B.\ K.\ wants to thank P.\ E.\ L.\ Rakow from DESY-IfH Zeuthen for his
support during the preparation of the diploma thesis. He initialized
the investigation of the latent heat.

\end{document}